\documentclass[twocolumn]{aastex631}
\usepackage{CJKutf8}
\usepackage{graphicx}

\usepackage{amsmath}
\usepackage{amssymb}
\usepackage{bm}

\shortauthors{Zhou et al.}
\graphicspath{{./}{figures/}}

\begin{document}

\title{Turbulent Transport of 
Dust Particles in Protostellar Disks: The Effect of Upstream Diffusion}


\author{Tingtao Zhou\begin{CJK*}{UTF8}{bsmi}(周廷弢)\end{CJK*}}
\affiliation{Department of Mechanical and Civil Engineering, California Institute of Technology, Pasadena, CA 91125, USA}
\author{Hong-Ping Deng\begin{CJK*}{UTF8}{gbsn}(邓洪平)\end{CJK*}}
\affiliation{Shanghai Astronomical Observatory, Chinese Academy of Sciences, Shanghai 200030, China}
\author{Yi-Xian Chen\begin{CJK*}{UTF8}{gbsn}(陈逸贤)\end{CJK*}}
\affiliation{Department of Astrophysics, Princeton University} 
\author{Douglas N. C. Lin\begin{CJK*}{UTF8}{bsmi}(林潮)\end{CJK*}}
\affiliation{Department of Astronomy \& Astrophysics, University of California, Santa Cruz, CA 95064, USA}
\affiliation{Institute for Advanced Studies, Tsinghua University, Beijing 100086, China}

\begin{abstract}

We study the long-term radial transport of micron to mm-size grain in protostellar disks (PSDs) based on diffusion and viscosity coefficients measured from 3D global stratified-disk simulations with a Lagrangian hydrodynamic method. While gas-drag tend to transport dust species radially inwards, stochastic diffusion can spread a considerable fraction of dust radially outwards (upstream) depending on the nature of turbulence.
In gravitationally unstable disks, we measure a high radial diffusion coefficient $D_{\rm r} \sim  H^2 \Omega$ with little dependence on altitude. This leads to strong and vertically homogeneous upstream diffusion in early PSDs.
In the solar nebula, the robust
upstream diffusion of $\mu$m to mm size grains not only efficiently transports highly refractory $\mu$m-size grains (such as those identified in the samples of comet 81P/Wild 2) from their regions of formation
inside the snow line out to the Kuiper Belt, but can also spread mm-size CAI formed in the stellar proximity to distances where they can be assimilated into chondritic meteorites.
In disks dominated by magnetorotational instability (MRI), the upstream diffusion effect is generally milder, with a separating feature due to diffusion being stronger in the surface layer than the midplane. This variation becomes much more pronounced if we additionally consider a quiescent midplane with lower turbulence and larger characteristic dust size due to non-ideal MHD effects. This segregation scenario helps to account for dichotomy of two dust populations' spatial distribution as observed in scattered light and ALMA images. 

 
\end{abstract}


\section{Introduction} \label{sec:intro}

The omnipresence of protostellar disks (PSDs) \citep{hartmann1998} provides birth sites and incubators for 
emergence of planets \citep{cameron1978, hayashi1985, lin1985, ida2004ApJ}.
The temperature distribution in these disks inferred from their observed spectral energy distribution
\citep{adams1987, hartmann1998}  matches well with those from theoretical thermal equilibrium models where 
the cooling rate from the disk surface is balanced by both intrinsic viscous dissipation and stellar 
irradiation \citep{chiang1997, garaud2007}.  This provides a natural division for the condensation sequence 
of terrestrial, asteroids, gas and ice giant planets \citep{ruden1986}.

In such a setting, refractory and volatile grains are expected to condense in the stellar proximity 
(within $\sim 0.1$ au) and outside the snow line (several aus) respectively. 
Yet, $\mu$m-size crystalline refractory 
grains are found in abundance among the Stardust return samples from comet 81P/Wild 2 \citep{brownlee2006}.
Their presence is also inferred from the IR spectra of other comets \citep{wooden2007} as well as 
extended regions (out to $\sim 10$ au) of PSDs \citep{bouwman2003, bouwman2008, vanboekel2005, 
watson2009}.

The crystalline structure requires the melt-down and re-condensation of amorphous silicate grains at 
{temperatures $\gtrsim 1000$K, depending on composition \citep{Bockel2002}}, while their implantation into comets occurs below the icy condensation
temperature $\lesssim 160$K. Therefore, very efficient mechanisms need to be invoked for these grains to be transported outwards after re-condensation in the stellar proximity. Explanations for this thermal paradox include large scale diffusion
\citep{morfill1984, clarke1988}, spreading of the solar nebula \citep{safronov1985, ciesla2007}, 
meridional circulation \citep{urpin1984, kley1992}, radiation-pressure driven outflow in the disk surface
\citep{takeuchi2003}, fluctuating X-wind \citep{shu2000, shu2001, 
shang2000}, and shock heating \citep{weidenschilling1998, wooden2007, desch2012, gong2019}. 

Optical and NIR maps of scattered light also indicate that in certain PSDs, the micron-size particles have larger radial
extent and thickness than those of larger mm-size particles inferred from ALMA images of reprocessed radiation \citep{Villenave2020,Benisty2022}. {This suggests that dust of different characteristic size, occupying different vertical layers of the disk, may have distinct capabilities of radial transport}.

In this paper, we revisit the diffusion hypothesis. The structure  and evolution of PSDs including the solar nebula are regulated by turbulent viscosity. Some causes of 
turbulence include the magneto-rotational instability (MRI) \citep{balbus1991} and gravitational instability (GI)
\citep{safronov1960, Toomre1964}  in non-magnetised \citep[see, e.g.,][]{Durisen2007} 
and magnetised disks \citep{Riols2019,Deng2020}. The chaotic turbulent motion
is correlated and leads to Reynold and Maxwell stress and effective
torque which transport gas angular momentum outwards, while inducing mass spreading
\citep{pringle1981,balbus1998}  and diffusion of passive contaminants including small grains
\citep{morfill1984, clarke1988} \footnote{An analogous turbulent-diffusion mechanism has also been invoked to account for the
shallow abundance gradient in disk galaxies \citep{clarke1989, yang2012, 
krumholz2018}.}  A key issue is whether turbulent diffusion of crystalline grains or
heavy-element tracers can diffuse against the inward gas flow 
and spread widely 
over extended regions of the disk.  We present in \S\ref{sec:simulations} three sets of
numerical simulation for MRI, GI
in the absence of magnetic  fields (GI-HD), and GI of disks
with magnetic fields (GI-MHD). 
Based on measurements from state-of-the-art simulations in \S\ref{sec:DustDiffusion}, we construct simplified 1D models of layered accretion disks and examine the global diffusion process of dust elements in turbulent disks in \S\ref{sec:dusttransport}, focusing on the evolution of their concentration distribution.
In \S\ref{sec:summary}, we summarize our results and discuss their implications.

\section{Simulations and analysis}\label{sec:simulations}

The simulations of GI-HD, MRI and GI-MHD turbulence are drawn from \cite{Deng2020} {dedicated to studying the interplay between magnetic fields and spiral density waves}. The simulations were carried out with a Godunov-type Lagrangian method, the meshless finite mass (MFM) method in {the GIZMO code} \footnote{The code is publicly available at \url{http://www.tapir.caltech.edu/~phopkins/Site/GIZMO.html}.} \citep{Hopkins2015} which shows excellent conservation property in GI disk simulations \citep{Deng2017}.  At sufficiently high resolution, MFM can also capture the MRI and the ensuing turbulence showing good agreement with grid-code simulations \citep{Deng2019,Deng2020}. {In all the simulations, the disk mass is about 0.07 the stellar mass and the disk is resolved by $>35$M particles/elements, i.e., a Jupiter mass object is resolved by half a million computational elements, among the highest resolution GI disk simulations.\footnote{We note that thanks to the adaptive nature of Lagrangian methods, even a 1M particle simulation produces converged results  comparable to some ultra-high resolution grid-code simulations (up to 2 billion effective cells) in GI disk modeling. See the Wengen  code comparison project test4, including one Lagrangian code, Gasoline and four grid codes (some of which with mesh refinement), Enzo, FLASH, CHYMERA and IU code at \url{http://users.camk.edu.pl/gawrysz/test4/}.}}
The initial surface density is inversely proportional to the radius spanning 5--25 au. The code units are 1 au, 1 solar mass and $G=1$.

\begin{figure*}[ht!]
\epsscale{1.2}
\plotone{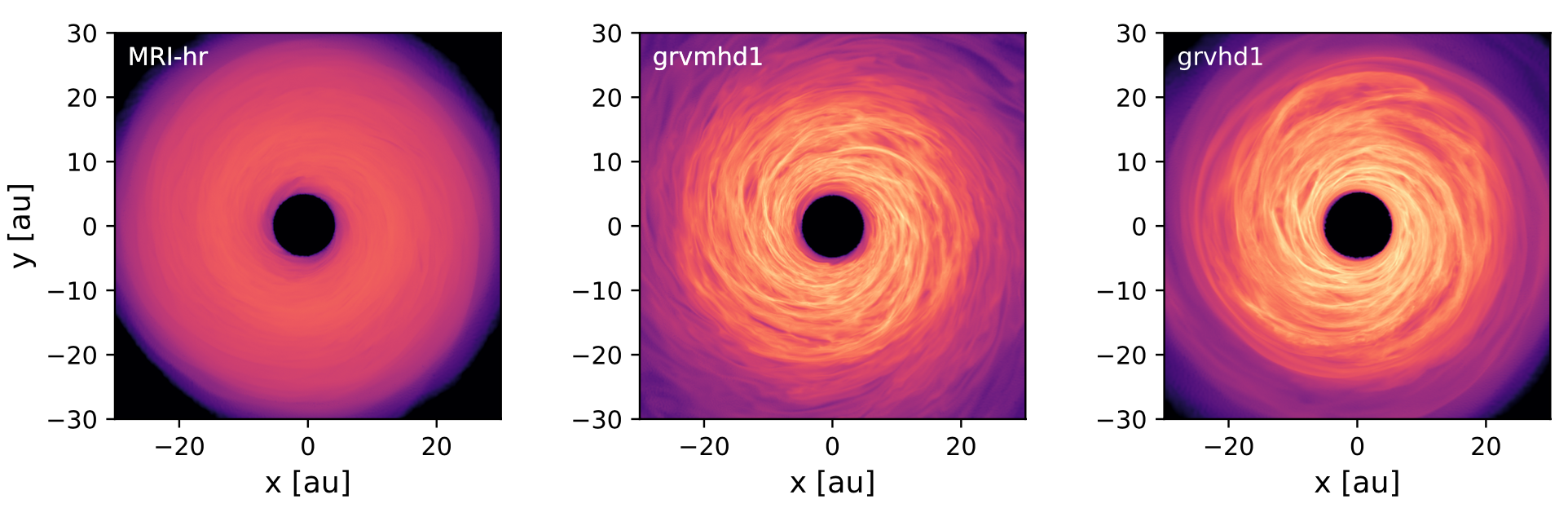}
\caption{The mid-plane volume density in quasi-equilibrium turbulence of MRI, GI-MHD and GI-HD simulations. The colors, from black to yellow, show the density ranging from $10^{-15}$ to $10^{-9}$ g/cm$^{-3}$ in a logarithmic scale.\label{fig:rho} }
\end{figure*}

In the following analysis, the MRI and GI-MHD turbulence corresponds to the model MRI-hr and grvmhd1 simulation. We then turn off the magnetic fields in the grvmhd1 model and run it for 400 additional years and this simulation is named grvhd1 here. We show representative mid-plane density maps \footnote{This is similar to the mid-row of figure 3 in \cite{Deng2020} and the high density inner edge in that figure is due to artificial effects during plotting.} in the quasi-steady state in Figure \ref{fig:rho}.  The MRI turbulence is mild with small density contrast while GI turbulence, regardless of the presence of magnetic fields, show strong density fluctuations. The flocculent spirals in the grvmhd1 model become more coherent after turning off the magnetic fields and saturating to grvhd1 resembling previous hydrodynamic simulations of gravitational instability \citep{Deng2017,Bethune2021}. This in turn confirmed the important back reaction of magnetic fields on the spiral density wave morphology \citep[for details, see][]{Deng2020}.

In the well-coupled regime, dusts are tied to the Lagrangian fluid elements so that we can directly analyse {the fluid elements' trajectories to uncover particle motions. This is different from injecting and analysing tracers of different sizes into either Lagrangian \citep{Rice2004} or Eulerian hydrodynamic simulations \citep{Haghighipour2003,Boley2010,Zhu2015,Riols2020,Baehr2021,Hu2021}.}
To our knowledge, this work amounts to be the first study of fluid tracers in global Lagrangian simulations of magnetised/self-gravitating accretion discs.

\begin{figure}[ht!]
\epsscale{1.2}
\plotone{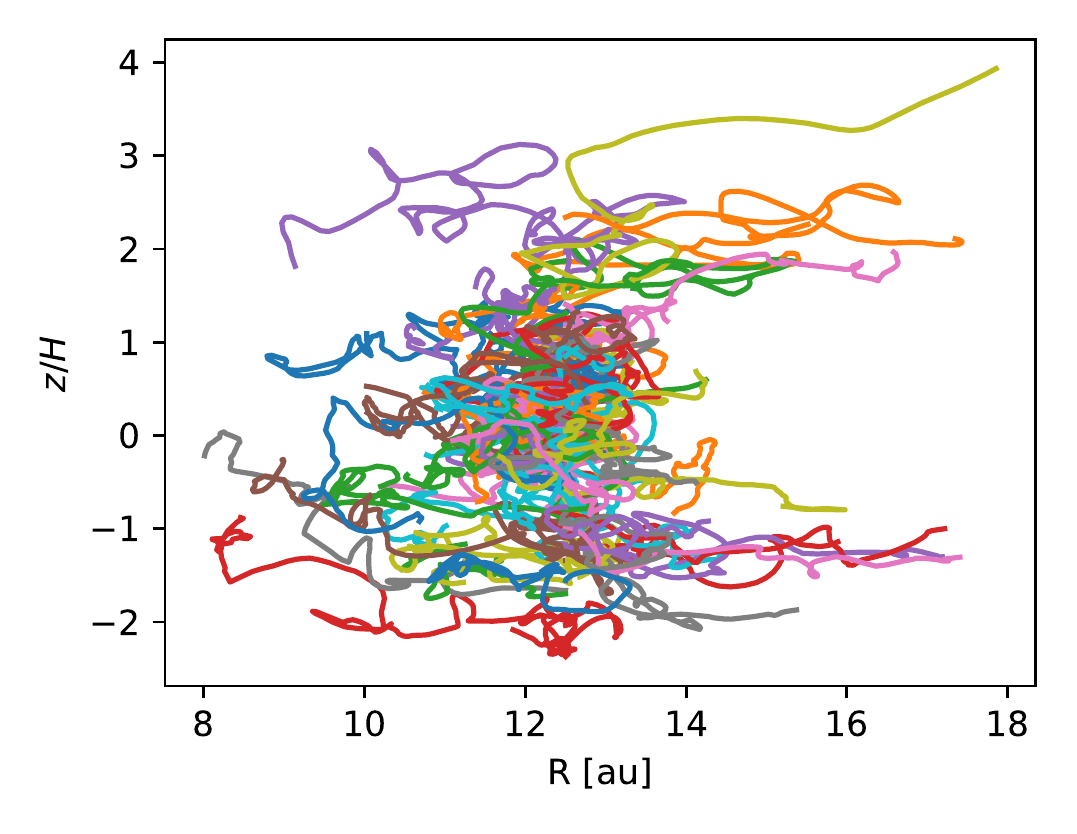}
\caption{The trajectories of particles in the MRI simulation ($R-z$ plane) initially at 12.5 au within five local orbits.\label{fig:trace} }
\end{figure}

We trace fluid elements in the quasi-equilibrium turbulence for 250 years (two rotation periods at 25 au), which is sufficient to identify the particles diffusion coefficients. We choose particles in small annuli centered at 10 au and 12.5 au with a width of 0.001 au. In each annulus, there are more than 2000 particles.  For example, we follow the trajectories of fluid elements in a narrow annulus of width 0.001 au in the MRI-hr simulation and plot the trajectories of 100 randomly chosen particles out of the $>$2000 particles in Figure \ref{fig:trace}. 

The three simulations have distinct thermophysics and vertical structures \citep[see, Figure 2 of][]{Deng2020} and thus, for ease in comparison, we need to normalise the statistics to an appropriate length scale. Specifically, the MRI-hr simulation is adiabatic and the accretion heats up the disk gradually; the grvmhd1 and grvhd1 saturates to a state where the heating is balanced by a prescribed  cooling at a timescale of the orbital period \citep{Deng2020}. In addition, the disk vertical structure is strongly fluctuating the GI simulations so that the traditional definition of the vertical scale-height is not adequate.

We adopt the dynamic scale-height ($H$) definition of  \cite{Ogilvie2018} where $H$ is the square root of the second vertical moment of the density field, i.e.,
\begin{equation}
H=\sqrt{\frac{\int \rho z^2  dz}{\int \rho dz}}.
\label{eq:h2}
\end{equation}

\begin{figure}[ht!]
\epsscale{1.2}
\plotone{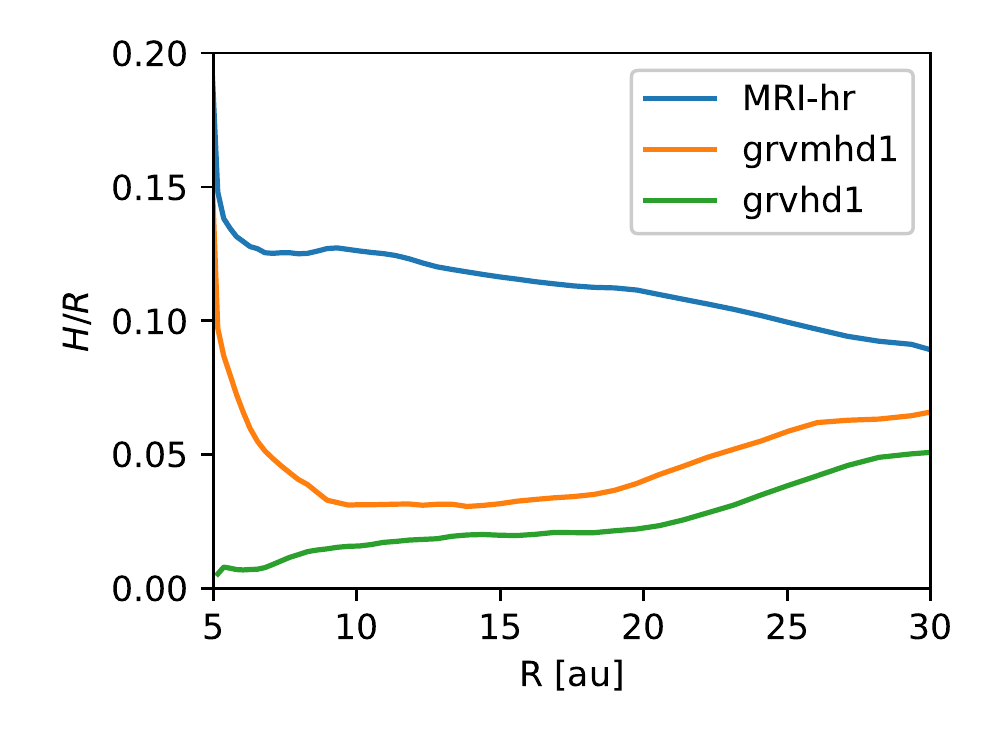}
\caption{The averaged dynamic scale height in the quasi-equilibrium turbulence\label{fig:h}. }
\end{figure}

It naturally reduce to the ratio between the speed of the sound ($c_s$) and the local rotational frequency ($\Omega$) for a vertical Gaussian profile. We take the time (250 years) and azimuthal average the dynamic scale-heights for the three simulations and show them in Figure \ref{fig:h}.

The adiabatic MRI-hr simulation is hot with a typical $H/R=0.1$ (due to the initial condition). The inner disc is relatively hotter due to a faster development (small $\Omega^{-1}$) of MRI turbulence than the outer region in the adiabatic simulation. The grvmhd1 and grvhd1 have significantly thinner/colder discs with $H/R$ around 0.03 and 0.02. In further analysis, we normalise displacements to $H$ to enable comparison across different models and with previous grid-code simulations \citep{Zhu2015,Riols2020}. However, for simplicity, a thermal profile with constant $H/R$ is adopted for GI, GI-MHD and MRI in the simplified 1-D dust transport model in \S \ref{sec:dusttransport}.

Runaway types of super-diffusion processes have been reported for certain highly supersonic 
turbulence, in which there is no well-defined diffusion coefficient \citep{colbrook2017}.
In contrast, our global simulations corroborate with previous local simulations 
\citep{Zhu2015,Shi2016} to show that GI and MRI, as turbulence typically on the 
sonic level (characteristic length scale comparable to $H$), induce regular 
Gaussian/Brownian diffusion of passive tracers. We can measure diffusion coefficients 
through the displacement of these tracers (see Figure \ref{fig:trace}), as elaborated in \S \ref{sec:DustDiffusion}. 
{Our approach is similar to the method applied in grid-code simulations to measure the 
diffusion coefficients of Lagrangian particles with finite Stokes number $T_S$.
As we mentioned, since particles represent gas itself in our simulations, we directly measure the 
diffusion coefficients of passive particles well-coupled with gas (the limit of small $T_S \rightarrow 0$), however, diffusivity measured in this limit are proven to be valid for small dust Stokes number up to $T_S\lesssim 0.1$ \citep{Zhu2015,Shi2016}, consistent with analytical scaling from \citet{Youdin2007}. }

The measured relevant 
diffusion coefficient discussed below are summarised in Table \ref{t:coeff}. Although $D \sim \alpha H^2\Omega$ is often applied in grid-based dust-gas hydrodynamic simulations \citep{Zhu2011,zhu2012dust,Chen2021}, we show that this approximation can be off by up to an order of magnitude. $D_r$ is the overall radial diffusion coefficient and $D_z$ is the overall vertical diffusion coefficient. To explore possibility of layered radial diffusion, we also divide disks into the disk midplane within $|z| < H$ and the disk atmosphere $|z| > H$, and measure the radial diffusion coefficients $D_{\rm mid}$ and $D_{\rm sur}$ for each layer. We generally expect that with a large $D_z$, the vertical mixing of dust/gas at different altitudes becomes very efficient such that $D_{\rm mid} \sim D_{\rm sur}$ whereas they may differ more significantly when $D_z$ is small. 

\begin{deluxetable}{c|c|c|c|c|c|c}[ht!]
  \tablenum{1}
  \tablecaption{dust diffusion coefficients  \label{t:coeff}}
  \tablewidth{\textwidth}
  \tablehead{
    \colhead{Run } & \colhead{Physics}& \colhead{$D_r$}& \colhead{$D_{\rm mid}$} & \colhead{$D_{\rm sur}$} & \colhead{$D_z$}& \colhead{$\alpha$} }
  \startdata
  grvhd1    &  GI      & 2.& 1.9  &2.1             &0.2 & 0.1 \\
    grvmhd1       &  GI-MHD    & 0.5  & 0.49  &  0.5  & 0.1     &  0.23  \\
     MRI-hr      & MHD     & 0.045& 0.033 & 0.07       & 0.013 & 0.02 \\ 
  \enddata
  \caption{
  The diffusion coefficients (given in a unit of $H^2\Omega$) by linear regression of the dust mean squared displacement (MSD). Note, for self-gravitating disk, only the early stage vertical MSD evolution reflects the fast vertical motion (see text). Here $D_{\rm mid}$ and $D_{\rm sur}$ are measured dust radial diffusion coefficients for particles with initial position $|z|<H$ and $|z|>H$, respectively. In the unit of $H^2\Omega$, $\nu$ corresponds to the value of $\alpha$ in the widely adopted $\alpha$ prescription for effective viscosity \cite{shakura1973}, and $D/\nu$ are referred to as Schmidt numbers. We see that while radial $D_r/\nu$ span a wide range, the vertical $D_z/\nu$ are always of order-unity. Schmidt numbers are also sometimes used to refer to the ratio between dust diffusivity and gas diffusivity, which we assume to be 1 in this study for small $T_S$ \citep{Youdin2007}. We will not be using this term in the rest of this paper to avoid confusion.
  } 
  \label{table:diffusion coeffs}
\end{deluxetable}

\section{Dust diffusion in real turbulence}
\label{sec:DustDiffusion}
\subsection{gravitationally unstable disks}

Young circumstellar disks are relatively massive to their hosts and can be unstable due to gas self-gravity forming spiral density waves \citep[see, e.g. reviews by][]{Durisen2007,Kratter2016}. The spiral density waves are critical for angular momentum transport and thus disk accretion. 
The spiral density wave also excites strong vertical circulation and likely powers a larges scale dynamo amplifying any weak magnetic fields to near thermal strength \citep{Riols2019,Deng2020,Lohnert2022combined}. The strong fields facilitates accretion in early discs significantly and may impact the population of planet formed via disk instability \citep{Deng2021}.

\begin{figure*}[ht!]
\epsscale{1.2}
\plotone{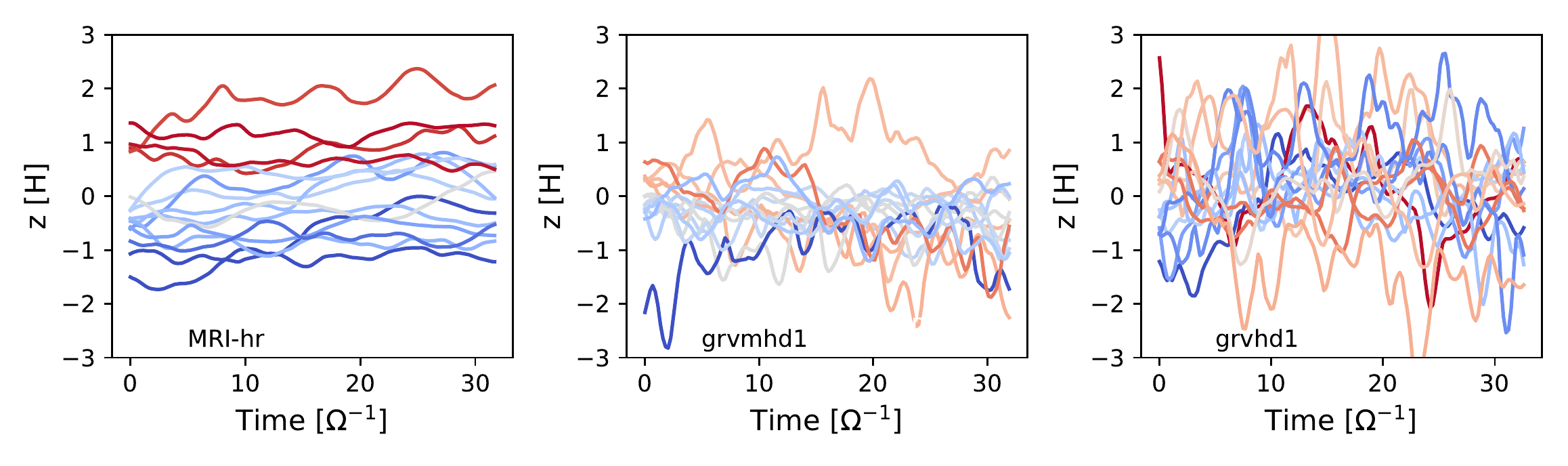}
\caption{The vertical motion of randomly chosen particles in the annuls centered at 12.5 au.  \label{fig:tracez}. }
\end{figure*}

In an gravitational unstable disks, the vertical circulation \citep{Boley2006,Deng2020}  besides spirals
efficiently mixes small particles at different altitudes above the disk mid-plane in local shearing box simulations \citep{Riols2020,Baehr2021}. In order to
characterize the difference in turbulent diffusion in disks regulated by MRI and GI, we plot the vertical motion of 15 randomly chosen particles in the annulus centered at 12.5 au in Figure \ref{fig:tracez}. Particles in the MRI turbulence show minor vertical motion (model MRI-hr), 
impliying a slow vertical diffusion. This weak coupling also enables particles in the disk 
mid-plane and surface to diffuse at different rates.  In contrast, the vertical motion 
over a dynamical time scale extend to the scale of $H$ in GI turbulence (models grvmhd1 and grvhd1). This random motion leads to efficient particle exchanges in the vertical direction and inhibits a significant differential radial diffusion between the disk mid-plane and surface.

\begin{figure}[ht!]
\epsscale{1.}
\plotone{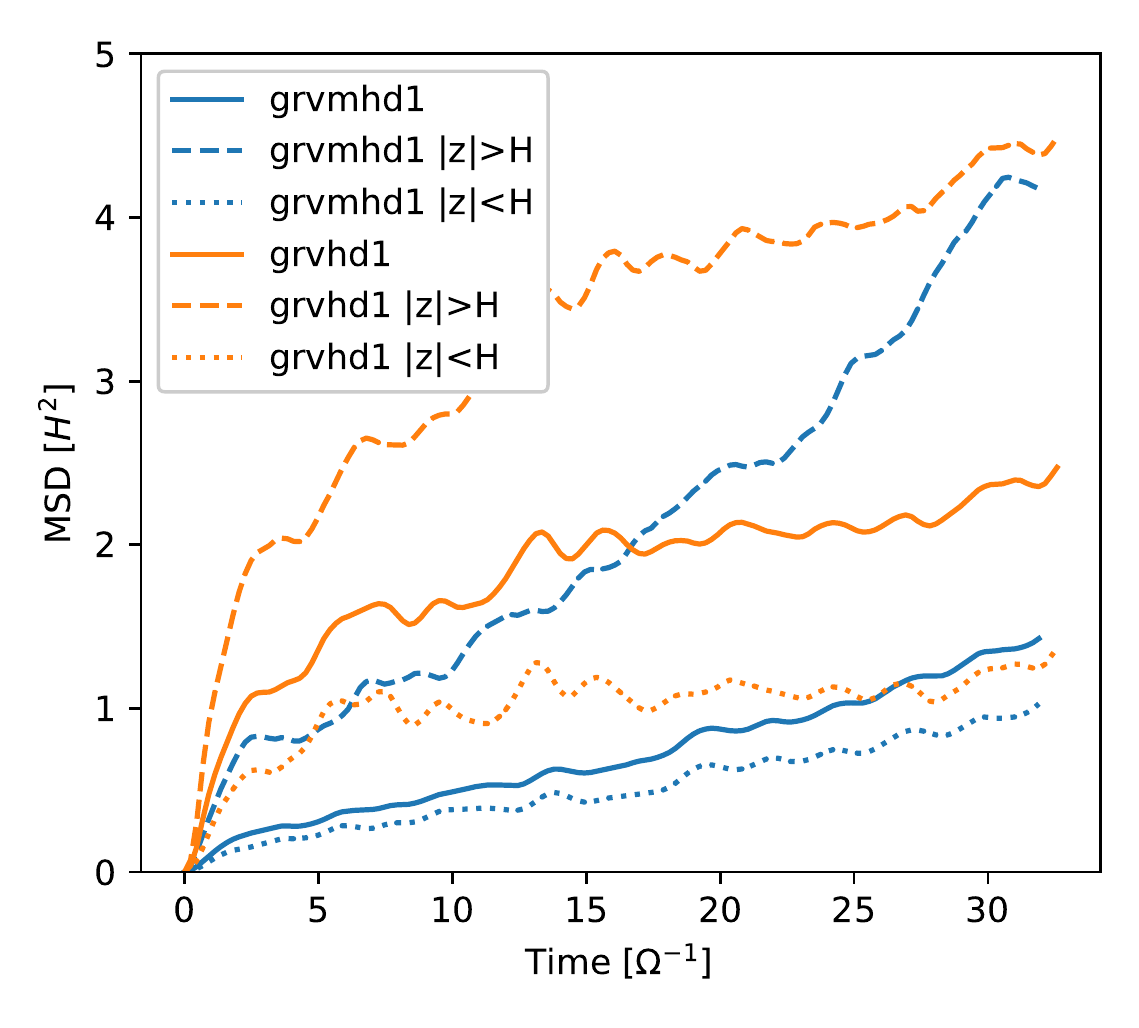}
\caption{The evolution (given in units of the local dynamical time) of the mean squared displacement along the vertical direction for particles in annuli centered at 12.5 au initially.   Here  the MSD  is  normalised  to  the dynamical scale-height at 12.5 au.   \label{fig:MSDz}. }
\end{figure}

The vertical diffusion process in grvmhd1 and grvhd1 can be quantified by the vertical mean-squared displacement (MSD).  In Figure \ref{fig:MSDz}, we plot the averaged vertical MSD for particles in the disk mid-plane ($|z|<H$) and surface ($|z|>H$) versus the all-particle average  MSD around 12.5 au, as functions of diffusing time. In both grvmhd1 and grvhd1, the vertical MSD experience a fast increase followed by a slower but steady growth. We measure a mean vertical diffusion rate of $D_z = 0.2H^2\Omega$ and  $0.1H^2\Omega$ for the grvhd1 and grvmhd1, comparable to their viscosity $\nu =\alpha H^2\Omega =  0.1H^2\Omega$ and $0.23H^2\Omega$ \citep{Deng2020}, also summarised in Table \ref{t:coeff}. After the particles in the narrow annulus (initially confined) spreads out along a large scale eddy, the growth of vertical MSD become slow in grvmhd1 or negligible in grvhd1. 
{The initial fast increase in MSD is typical for high altitude particles ($|z|>H$), suggesting faster flow in the disk surface layer (dashed lines). However, over a few dynamical timescales the MSD growth quickly saturate towards a lower pace that converges with the average diffusion rate (dotted lines), since particles’ movement is hindered by their rapid circulation across lower altitudes ($|z|\sim H$), where diffusion is much slower.
In principle, we can measure somewhat different vertical diffusion coefficients $D_{z, mid} \ll D_{z, sur}$ from the initial trend of dashed and dotted lines in Figure \ref{fig:MSDz}. 
Nevertheless, since vertical mixing is efficient suggesting particles commute across different layers frequently (as shown in Figure \ref{fig:tracez}), it would be more meaningful to simply apply the average diffusion coefficient in following discussions for disks with gravitational instability.}

\begin{figure}[ht!]
\epsscale{1.}
\plotone{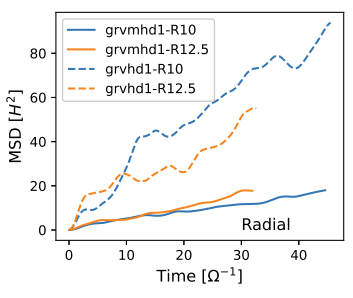}
\caption{The evolution (given in units of the local dynamical time) of the mean squared displacement for particles.  Here  MSD  is  normalised  to  the dynamical scale-height at 12.5 au or 10 au.   \label{fig:MSDgrv}. }
\end{figure}

Moreover, due to efficient vertical mixing, the radial MSD shows little altitude dependence $D_r\approx D_{\rm mid}\approx D_{\rm sur}$. As a result, we only plot the average all-particle MSD (used to measure $D_r$) for all particles centered at 10 au and 12.5 au in Figure \ref{fig:MSDgrv}. At the two different radii, the MSD evolves similarly so that our results apply to general GI and GI-MHD turbulence. Similar to the vertical MSD evolution, the radial MSD also increase rapidly at the early stage when the particles are moved around by the vertical circulation. However, particles can hop onto new eddies and be radially transported over 
large distances. As a result, the radial MSD continues to grow. In the grvmhd1 model, the radial diffusion rate is $D_r = 0.5H^2\Omega$, twice the viscosity. In the grvhd1 model, the radial diffusion rate is $2H^2\Omega$ which is much larger than its viscosity $0.1 H^2\Omega$, as summarised in the first 2 rows of
Table \ref{t:coeff}. 

\subsection{MRI turbulence}

Magnetised accretion disks are prone to the MRI provided sufficient ionization \citep{balbus1991}, which further leads to vigorous turbulence that transports angular momentum outwards enabling gas accretion. In ideal MHD simulations, MRI turbulence has a typical effective viscosity $\nu=\alpha H^2\Omega$, with $\alpha\sim 0.02$ in grid-code \citep{Simon2012,Beckwith2011} as well as Lagrangian simulations \citep{Deng2019,Deng2020,Wissing2022}. However, realistic PSDs are poorly ionized so that only the surface layer may be MRI active and turbulent or the whole vertical extent is laminar with winds launched at high altitude \citep{gammie1996, Okuzumi2011modeling,Bai2013wind,Riols2020ring}. In that case, the constrast between
$D_{\rm sur}$ and $D_{\rm mid}$ would be even more pronounced.  Our ideal MHD (active 
at all altitudes) model is admittedly oversimplified and it offers the most conservative
illustration for the layered dust diffusion tendency in MRI turbulence.
\begin{figure}[ht!]
\epsscale{1.2}
\plotone{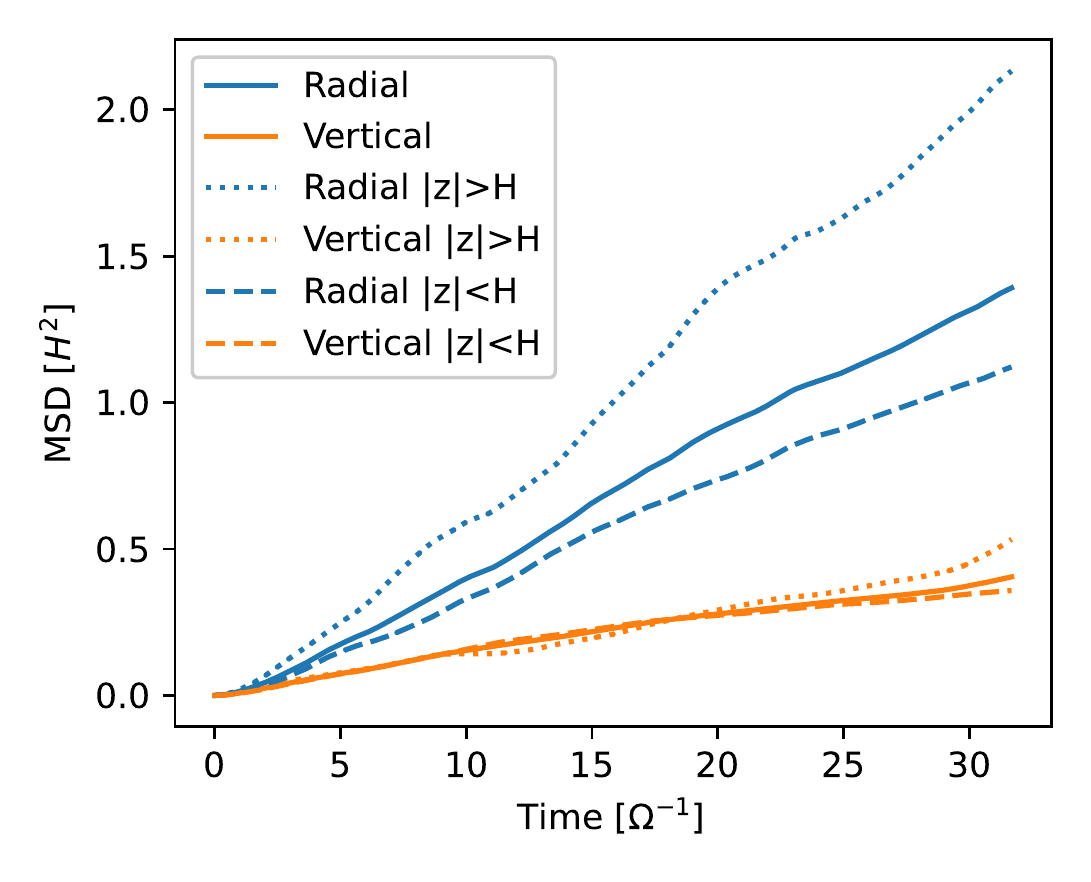}
\caption{The evolution (given in units of the local dynamical time) of the mean squared displacement for particles initially at 12.5 au for the MRI-hr model. Here MSD is normalised to the dynamical scale-height at 12.5 au. \label{fig:MSD125}. }
\end{figure}

In Figure \ref{fig:MSD125}, we plot the MSD evolution for all particles in the annulus centered at 12.5 au and particles initially in the disk body with $|z|<H$ and disk surface with $|z|>H$ separately. There are some 1700 and 600 particles in the disk body and surface respectively. By the end of the simulation about 200 particles crossed the $z=\pm H$ boundary.  If we restrict our analysis to those particles confined in the same (mid-plane or surface) portion of the disk \citep[see, e.g.][]{Zhu2015}, we would find almost identical results to those in Figure \ref{fig:MSD125}. The particles show a mild uniform vertical diffusion rate about $D_z = 0.013H^2\Omega$ which is smaller than the viscous $\alpha$ parameter in units of $H^2\Omega$. A similar analysis for the particles in an annulus centered at 10 au reveals a vertical diffusion rate about $0.017H^2\Omega$ close to $\alpha$. However, in Table \ref{t:coeff}, we restrict ourselves to the reference value measured at 12.5 au since the latter is less likely affected by the inner boundary (see Figure \ref{fig:h}).

In the MRI context, the diffusion rates start to show dependence on the vertical altitude. Particles in the disk mid-plane diffuse at a rate $D_{\rm mid} = 0.033H^2\Omega$ whereas those at the disk surface diffuse at a faster rate of $D_{\rm sur} = 0.07H^2\Omega$. This introduces a factor of two difference in their characteristic radial diffusion ``velocity" $\sim D_{\rm r}/R$. The analysis for particles centered at 10 au show similar radial diffusion rates within a few percent deviation. 

Since both $\alpha$  and $D_z \approx \alpha$ are small, vertical mixing in the MRI-prone disks is much less efficient than the GI cases. Under such circumstance, the considerable difference between
$D_{\rm mid} < D_{\rm sur}$ 
introduces potentials for a layered diffusion process. Moreover, to go beyond our ideal MHD simulation, in realistic disks the moderate turbulence rendering $\alpha$ and $D_{\rm sur}$ as measured in our simulations may only be active in the surface layer \citep{gammie1996}, while residue $\alpha$ and $D_{\rm mid}$ is much smaller in the dead zone below the atmosphere.

To study the long-term evolution of grain in fore-mentioned environments, we construct a 1D two-fluid disk model to further explore the long-term implications of the
diffusion process with or without layered structure. 



\section{Dust transport models}
\label{sec:dusttransport}

In the case of GI or GI-MHD turbulence, the radial diffusion coefficients are similar and independent of the vertical position in the disk, as in Table~\ref{table:diffusion coeffs}. We write down the evolution for the dust concentration $C$, assuming cylindrical symmetry  so that there will be no explicit dependence on the azimuthal angle $\theta$.
\begin{align}
\Sigma \frac{\partial C}{\partial t}  = & \frac{1}{R}\frac{\partial}{\partial R}\left( R D \Sigma\frac{\partial C}{\partial R} \right) \nonumber \\
&- \frac{1}{R}\frac{\partial}{\partial R}\left( R \Sigma v_{\rm d} C \right)
+ C \frac{1}{R}\frac{\partial}{\partial R}\left(R\Sigma v_g\right)\label{eqn:dust-1layer}\\
\frac{\partial \Sigma}{\partial t}  = & \frac{3}{R} \frac{\partial}{\partial R}\left( R^{1/2} \frac{\partial}{\partial R} ( \Sigma \nu R^{1/2}) \right)\label{eqn:gas density evo} 
\end{align}

where $\Sigma$ is the gas density, $C$ is the dust-to-gas surface density ratio, and Equation \ref{eqn:gas density evo} describes the evolution of the gas density~\citep{ruden1986}.
In Equation \ref{eqn:dust-1layer} 
we generalize the formulation of \citet{morfill1984,clarke1988}, where we include the radial velocities of dust $v_d$, which could differ from the radial velocity of the gas $v_g$. We note that the dust back reaction is often ignored in the gas nebula evolution (Equation \ref{eqn:gas density evo}) given the low dust concentration level of a few percent \citep[see, e.g.,][]{Liu2022}. The gas advective velocity in the radial direction follows \citep{lyndenbell1974,Armitage2019} 

\begin{align}
v_g  =&-\frac{3}{\Sigma R^{1/2}}\frac{\partial}{\partial R}(\nu \Sigma R^{1/2})\\
=&-\frac{3\nu}{R} \left( \frac{\partial \ln\Sigma}{\partial \ln R } + \frac{\partial \ln \nu}{\partial \ln R} + \frac{1}{2} \right). 
\label{eqn:gas velocity}
\end{align}

For dust completely coupled with gas $v_d = v_g$, we can recover Equation 2.1.4 of \cite{clarke1988}. The motion of the dust particles are influenced by
the hydrodynamic drag by the disk gas. Considering dust particles 
with finite inertia, 
their radial velocity  
~\citep{Takeuchi2002,takeuchi2005attenuation,zhu2012dust} is
\begin{eqnarray}
v_d = \frac{v_g T_S^{-1} - \eta v_K}{T_S + 1/T_S}
\end{eqnarray}
where $\eta=- h^2 d\ln P/d\ln R$ with $h=H/R$ is a measure for the sub-Keplerian rotation of gas and $v_K = \sqrt{GM/R}$ is the Keplerian velocity.
The Stokes number, or dimensionless stopping time in the Epstein regime \citep[e.g.][]{Birnstiel2010}
\begin{equation}
\begin{aligned}
T_S &= \dfrac{\pi}{2}\frac{\rho_{d} s }{\Sigma} \\&= 0.0015 \times\left(\frac{s}{1 \mathrm{~mm}}\right)\left(\frac{\rho_{d}}{1 \mathrm{~g} \mathrm{~cm}^{-3}}\right)\left(\frac{100 \mathrm{~g} \mathrm{~cm}^{-2}}{\Sigma}\right),\label{eqn:Ts}
\end{aligned}
\end{equation}
depends on individual dust grains' radius $s$, internal \textit{physical} density $\rho_d$, 
and the gas volume density $\rho_g$. 

In the case of MRI turbulence, the diffusion coefficients measured for the middle plane and surface of the disk are different, and 
we correspondingly treat the dust distribution as two stratified layers, with the middle plane dust concentrations  $C_{\rm mid} (R,\theta,t) = \int_0 ^H \rho_d dz /\int_0 ^H \rho_g dz$ and surface layer concentration $C_{\rm sur} (R,\theta,t) =\int_H ^\infty \rho_d dz /\int_0 ^H \rho_g dz$ where $\rho_g$ is the gas density and $\rho_d$ 
is the dust density.  
Locally if the two layers have different densities, there would be a linear mass transfer rate $\lambda=D_{z}/H^2$ between the two layers. We can write down the coupled two-layer diffusion problem as

\begin{widetext}
\begin{align}
\phi\Sigma \frac{\partial C_{\rm mid}}{\partial t}  = & \frac{1}{R}\frac{\partial}{\partial R}\left( R D_{\rm mid} \phi\Sigma\frac{\partial C_{\rm mid}}{\partial R} \right) - \frac{1}{R}\frac{\partial}{\partial R}\left( R \phi \Sigma v_{\rm d,mid} C_{\rm mid} \right)
+ C_{\rm mid} \frac{1}{R}\frac{\partial}{\partial R}\left(R\phi\Sigma v_g\right)
+ F_z \label{eqn:dust-gas-fullcouple1}\\
(1-\phi)\Sigma \frac{\partial C_{\rm sur}}{\partial t}  = & \frac{1}{R}\frac{\partial}{\partial R}\left( R D_{\rm sur} (1-\phi)\Sigma\frac{\partial C_{\rm sur}}{\partial R} \right) -  \frac{1}{R}\frac{\partial}{\partial R}\left( R (1-\phi)\Sigma v_{\rm d,sur} C_{\rm sur} \right)
+ C_{\rm sur}\frac{1}{R}\frac{\partial}{\partial R}\left(R(1-\phi)\Sigma v_g\right)
- F_z
\label{eqn:dust-gas-fullcouple2}
\end{align}
\end{widetext}
where $\phi$ is the fraction of gas mass in the middle plane. {$F_z=\Sigma D_z/H^2((1-\phi)C_{\rm sur}-\phi C_{\rm mid})$ is the vertical diffusion term.} $v_{\rm d,mid}$ and 
$v_{\rm d,sur}$ are the  radial velocity of dust in the mid-plane and surface layer. 

\begin{figure*}[ht!]
\includegraphics[width=0.98\textwidth]{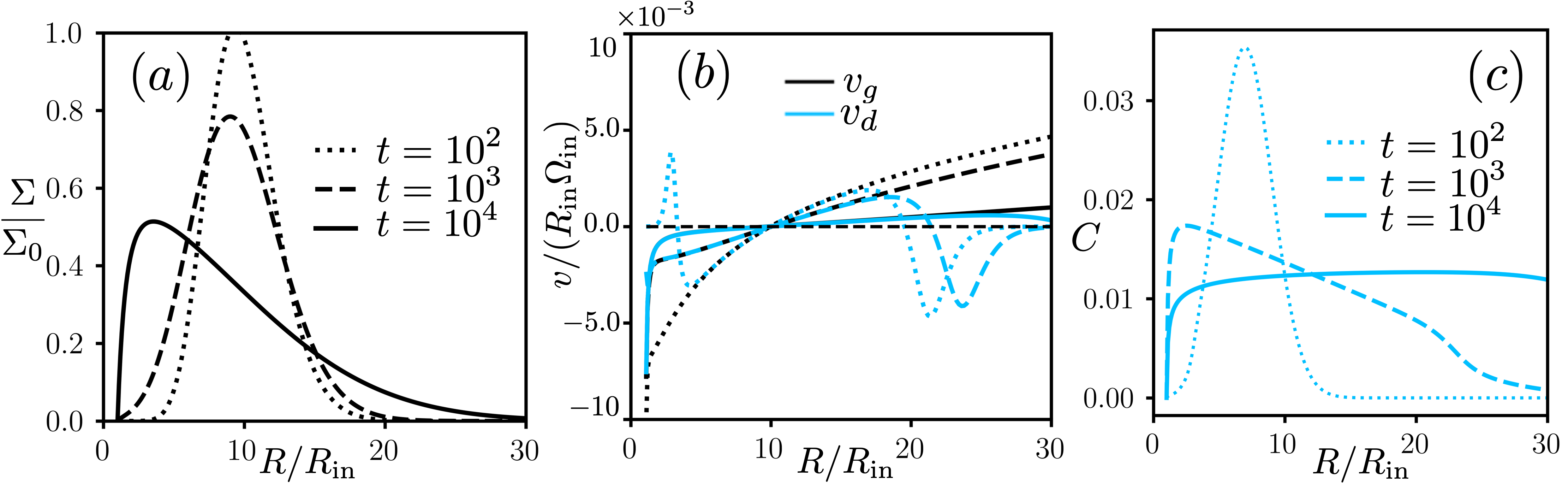}
\caption{
Coupled evolution of gas and dust in an early disk with GI turbulence. The gas density is initialized by a bell curve centered around $R_i=10$, and it evolves according to Eq.~\ref{eqn:gas density evo}, with an absorbing inner boundary.
$\alpha=0.1$, $D_{\rm r}=2~H^2\Omega$.
(a) The gas density evolution. 
(b) the gas drift velocity (black lines) and the dust velocities of the middle plane (blue lines) and the surface layer (red lines).
(c) The evolution of  dust concentration (blue lines).
In all three plots the times are indicated by linestyles: dotted for $t=100$, dashed for $t=1000$ and solid for $t=10000$.  
\label{fig:evolution-dust-gas-coupled-earlybump} }
\end{figure*}

For simplicity, we take $\phi = 0.7$ approximating a Gaussian vertical profile, assuming a constant $h=H/R = 0.05$, 
and therefore $\eta=(1-d\ln\Sigma d\ln R)h^2$ and
$\nu=\alpha h^2 \sqrt{GMR} \propto R^{0.5}$.

We choose different $\alpha $ for different turbulence models according to Table \ref{t:coeff} unless otherwise specified. 
The inner boundary is set at a normalizing length scale $R_{\rm in}=1.0$ with 
absorbing boundary condition for the dust concentration $C(R=R_{\rm in})=0$. For the GI-dominated disk as shown in Fig.~\ref{fig:evolution-dust-gas-coupled-earlybump} the inner boundary is absorbing for the gas density $\Sigma(R=R_{\rm in})=0$, while for the other cases presented below we use Dirichlet boundary condition for gas density (to maintain a steady power law). 
The outer boundary is set at $R_{\rm out}=130$ with a Dirichlet boundary condition for the gas density and absorbing condition for the dust concentration(s). 
The time unit is set to $\Omega_{\rm in}^{-1}$, the Keplerian dynamical timescale at $R_{\rm in}$. Focusing on sub-mm size to micron size grains, we assume a characteristic dust size corresponding to $T_S=0.001$, which is well-coupled with gas (see Equation \ref{eqn:Ts}) in most of our models with vertically-homogeneous $\alpha$, neglecting vertical differentiation of dust characteristic size due to sedimentation \citep{takeuchi2003,garaud2004}. This can be justified by considering the maximum dust Stokes number allowed by fragmentation \citep{Ormel2007,Birnstiel2010,Chen2020,Li2019}:

\begin{equation}
    T_{\rm S,max} \approx \dfrac{u_f^2}{2\alpha c_s^2},
    \label{eqn:frag}
\end{equation}

for fragmentation velocity $u_f\sim 1 $m/s, typical sound speed $c_s \sim 1$km/s and $\alpha > 0.02$, the characteristic dust species' size growth by coagulation will be halted at $T_S$ smaller than a few 0.001, even in the midplane. Because these dust species already couple fairly well with dust, the diffusion of $\mu$m crystalline grains with even smaller $T_s$ will be even better coupled to gas. We will discuss possible effects of having a dead zone with much smaller midplane turbulence in \S \ref{sec:MRI_with_dead_zone}, where characteristic dust species may be allowed to grow to larger than mm-size.

In the following sections we apply one- and two-layered diffusion model to PSDs at different stages of evolution with the dust diffusion parameter extracted from the turbulence simulations 
as summarised in Table \ref{t:coeff}. 

\begin{figure*}[ht!]
\includegraphics[width=0.98\textwidth]{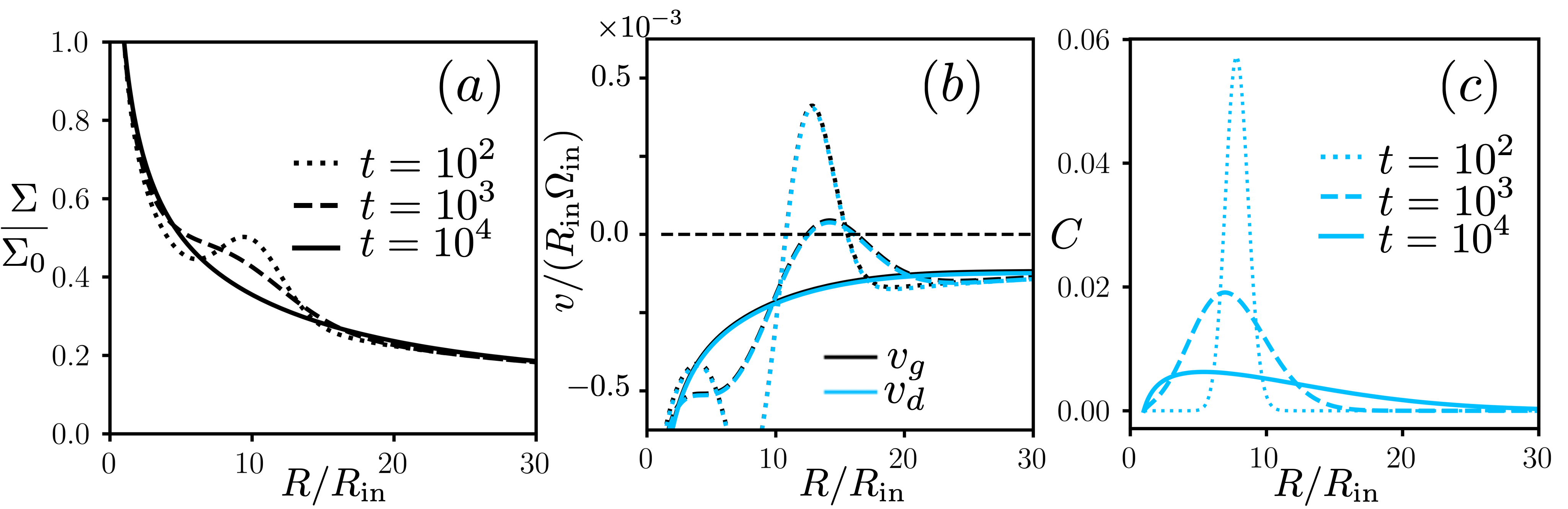}
\caption{
Coupled evolution of gas and dust in a GI-MHD disk perturbed from the steady-state power-law distribution.
$\alpha=0.2$, $D_{\rm r}=0.5~H^2\Omega$.
(a) The gas density evolution and (b) the gas drift velocity (black lines) and the dust velocity (blue lines) almost overlap throughout the evolution, due to small grain sizes with $T_S=0.001$.
(c) The evolution of  dust/gas ratio from solving Equations \ref{eqn:dust-1layer} and \ref{eqn:gas density evo}.
In all three plots the times are indicated by linestyles: dotted for $t=100$, dashed for $t=1000$ and solid for $t=10000$. \label{fig:evolution-dust-gas-coupled-grvmhd} }
\end{figure*}

\begin{figure*}[ht!]
\includegraphics[width=0.98\textwidth]{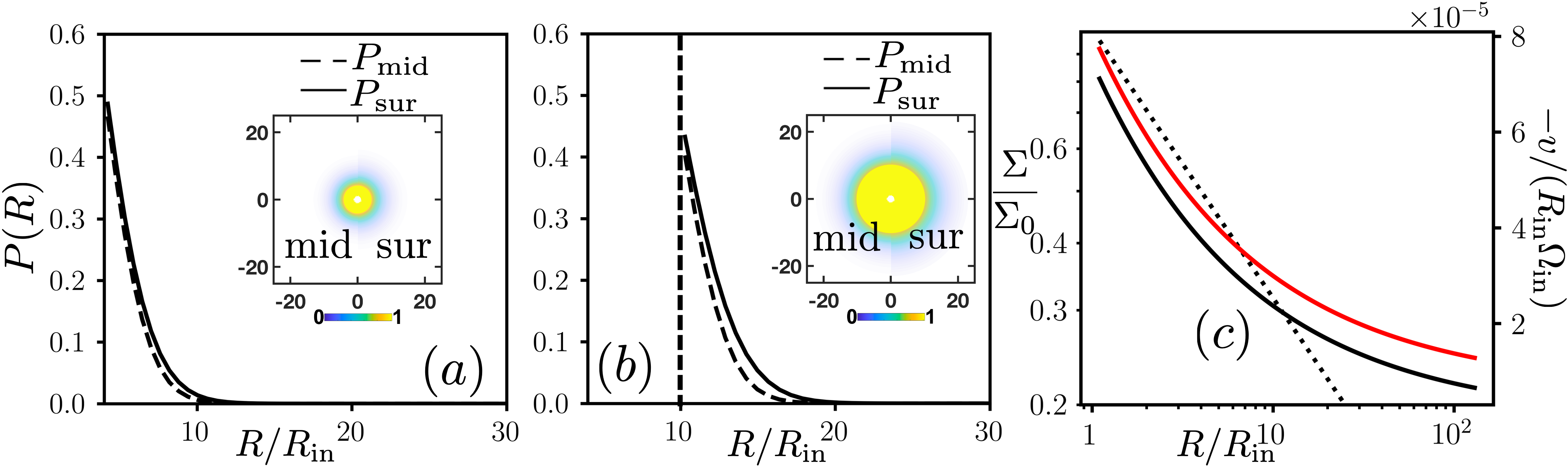}
\caption{The maximum upstream fractions $P(R)$ {\bf (dimensionless)} at $t=1\times10^4~\Omega_{\rm in}^{-1}$, due to ideal MRI turbulence, computed for the initial dust ring positioned at (a) $R_0=4$ and (b) $R_0=10$ with $\alpha=0.02$, $D_{\rm mid}=0.033~H^2\Omega$, $D_{\rm sur}=0.07~H^2\Omega$, $D_z=0.013~H^2\Omega$. $T_{\rm S,mid}=T_{\rm S,sur}=0.001$.
Solid lines represent the surface layer, and dashed lines are for the middle plane layer.
The inset shows the inferred disk concentration images, with the left half representing the middle plane layer and the right half representing the surface layer.
(c) The steady state power-law gas density (dotted black line, label on the left), gas radial velocity (black solid line) and the dust radial drag velocity (red line, label on the right).
\label{fig:steadystate-idealmhd}}
\end{figure*}

\begin{figure*}[ht!]
\includegraphics[width=0.98\textwidth]{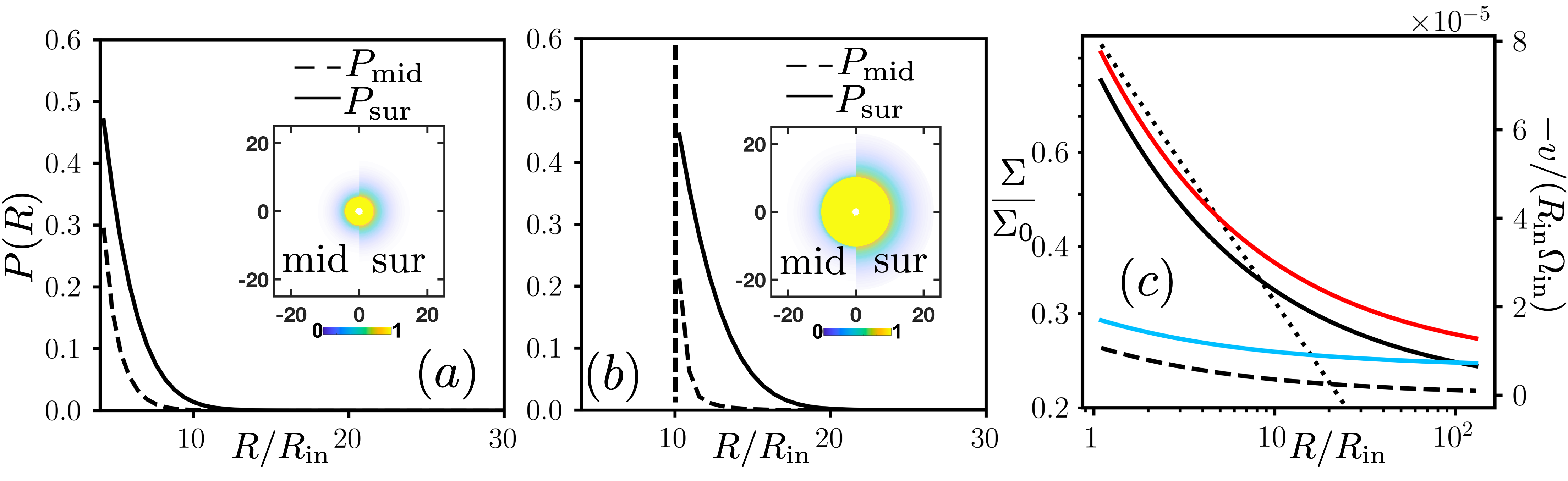}
\caption{The maximum upstream fractions $P(R)$ at $t=1\times10^4~\Omega_{\rm in}^{-1}$, due to non-ideal MRI turbulence, computed for the initial dust ring positioned at (a) $R_0=4$ and (b) $R_0=10$ with $\alpha_{\rm mid}=0.003$,  $\alpha_{\rm sur}=0.02$, $D_{\rm mid}=D_z=0.003~H^2\Omega$, $D_{\rm sur}=0.07~H^2\Omega$. $T_{\rm S,mid}=T_{\rm S,sur}=0.001$.
Solid lines represent the surface layer, and dashed lines are for the middle plane layer.
The inset shows the inferred disk concentration images, with the left half representing the middle plane layer and the right half representing the surface layer.
(c) The steady state power-law gas density (dotted black line), gas radial velocities of the middle plane (dashed black line) and the surface layer (solid black line), and the dust radial drag velocities of the middle plane (blue line) and the surface layer (red line). 
\label{fig:steadystate-nonidealmhd}}
\end{figure*}

\begin{figure*}[ht!]
\includegraphics[width=0.98\textwidth]{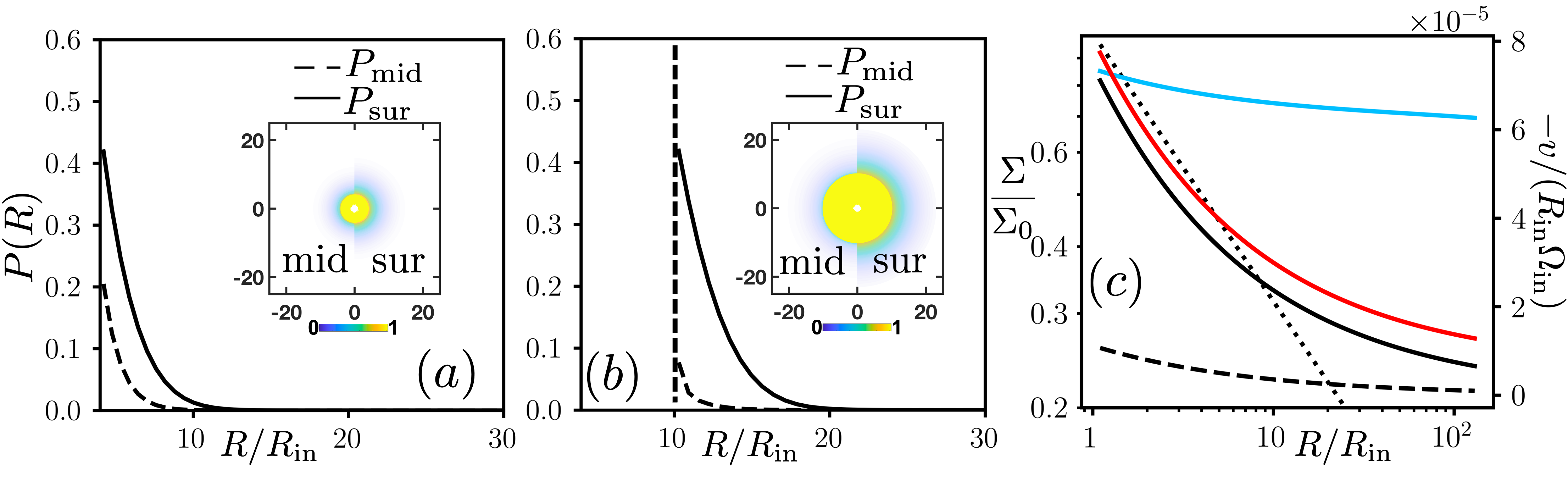}
\caption{The maximum upstream fractions $P(R)$ at $t=1\times10^4~\Omega_{\rm in}^{-1}$, due to non-ideal MRI turbulence and differential stopping times, computed for the initial dust ring positioned at (a) $R_0=4$ and (b) $R_0=10$ with $\alpha_{\rm sur}=0.02$, $\alpha_{\rm mid}=0.003$, $D_{\rm sur}=0.07~H^2\Omega$, $D_{\rm mid}=D_z=0.003~H^2\Omega$. The stopping times are chosen as $T_{\rm S,mid}=0.01,~T_{\rm S,sur}=0.001$.
Solid lines represent the surface layer, and dashed lines are for the middle plane layer.
The inset shows the inferred disk concentration images, with the left half representing the middle plane layer and the right half representing the surface layer.
(c) The steady state power-law gas density (dotted black line), gas radial velocities of the middle plane (dashed black line) and the surface layer (solid black line), and the dust radial drag velocities of the middle plane (blue line) and the surface layer (red line)..
\label{fig:steadystate-Ts}}
\end{figure*}

\subsection{Coupled dust-gas evolution in a gravitationally unstable disk}
\label{sec:GI_MHD}

As the PSD evolves in time, the disk mass and accretion rate gradually decline, regions of gravitational instability will recede and MRI turbulence will dominate in the inner disk \citep{Nakamoto1994,Hueso2005}.
We first examine the coupled evolution of dust and gas densities in the youngest GI-dominated 
disks \citep{2022ApJ...934..156X}, by solving Equations \ref{eqn:dust-1layer} and \ref{eqn:gas density evo} for a homogeneous layer of dust with $T_s = 0.001$, using $\alpha=0.1$ and $D_r = 2$.
Considering the early disk to be very dynamic and GI dominated, we demonstrate an initially bell-shaped gas density using a Gamma distribution, as shown in Figure \ref{fig:evolution-dust-gas-coupled-earlybump}(a)
\begin{eqnarray}
\Sigma(R) = \Sigma_0 \frac{6 (R-R_{\rm in})^{k-1} e^{-(R-R_{\rm in})/\theta}}{\Gamma(k)\theta^k}
\label{eqn:bell}
\end{eqnarray}
where $k=15$ and $\theta=0.6$ are the shape parameters and $\Sigma_0$ is the gas surface density unit.
The gas density evolution induces a separatrix around the peak of the initial distribution $R_{\rm peak}\sim 10$, where gas drift velocity is inward for $R<R_{\rm peak}$ and outward for $R>R_{\rm peak}$, as seen in Figure \ref{fig:evolution-dust-gas-coupled-earlybump}(b). The factor $6$ is chosen such that the peak value of gas density is $\approx \Sigma_0$.
For the initial dust density, we choose a narrow ring around an arbitrary radius $R_0=8$
analogous to the delta function adopted by \cite{clarke1988}. After {10000 dynamical timescales at the inner boundary}, 
significant outward/upstream diffusion of the dust concentration is seen in Figure \ref{fig:evolution-dust-gas-coupled-earlybump}(c). By the end of the simulation (solid lines), both the gas and dust density profiles have significantly flattened, while the gas and dust advection velocities have coupled with each other, conserving the location of the velocity separatrix. Due to large values of $D_r = 2.0$, a considerable portion of grains can diffuse out to $R\gg  R_0$ although its peak will still drift towards the inner boundary at velocities close to $v_g$, which is negative for $R\lesssim R_{\rm peak}$.

At a later stage, significant GI and MHD turbulence co-exist to influence disk evolution. Reading off Table \ref{table:diffusion coeffs}, we apply viscosity coefficient $\alpha=0.2$ and radial diffusion coefficient $D_r = 0.5$. We find that the Bell curve initial condition (Equation \ref{eqn:bell}) follows a similar diffusion process, only slower in time by a factor of 3-4 due to change in $D_r$. To mimic a relatively more stable disk profile, we then consider another initial gas density by superimposing a Gaussian bump on a steady-state power-law. 
The gas bump is also centered at around $R_{\rm peak}=10$. 
\begin{eqnarray}
\Sigma=\Sigma_0 R^{-0.5} + 0.2\Sigma_0 e^{-\frac{(R-R_{\rm peak})^2}{2\sigma_g^2}}
\end{eqnarray}
with the Gaussian spread chosen as $\sigma_g=2.0$, as shown in Figure.~\ref{fig:evolution-dust-gas-coupled-grvmhd}(a) with dotted lines. 
The gas/dust radial velocities are shown in Figure \ref{fig:evolution-dust-gas-coupled-grvmhd}(b), and the typical dust concentrations at times $t=100,1000,10000$ are shown in Figure \ref{fig:evolution-dust-gas-coupled-grvmhd}(c). Throughout the simulation, the dust and gas velocities are tightly coupled with each other. At the end of the simulation, the gas density relaxes towards a power law, and 
the dust concentration always spread out beyond the initial ring position
$R_0$, although its peak still migrates inward due to a generally negative advection velocity profile. {For GI-MHD turbulence we also tried to start with a smooth steady-state gas profile without the Gaussian bump (with $v_d \approx v_g$ being negative throughout the simulation), and found the evolution of dust concentration is very similar to Figure \ref{fig:evolution-dust-gas-coupled-grvmhd}(c), confirming that outward radial transport is dominated by diffusion rather than advection as $D_r > \alpha$.}


\subsection{Upstream diffusion of dust in a steady-state MRI disk with ideal MHD turbulence}
\label{sec:idea_MHD}

For quasi-steady MRI turbulence applicable to a later stage of PSD evolution, in the simulation we measure a radial accretion flow with a velocity $\sim -10^{-4} v_K$. With $H/R \lesssim 0.1$ and $\alpha \sim
2 \times 10^{-2}$, this value is on the order of the radial velocity in the steady solution \citep[e.g.][]{clarke1988}

\begin{equation}
v_g=-\frac{3\nu}{2R}.
\end{equation}

Here we examine dust evolution also based on the steady-state model for gas accretion flow.
We demonstrate the possibility of upstream diffusion with the corresponding steady-state $\Sigma =\Sigma_0 R^{-0.5}$ profile 
. Since the gas evolution equation gives $\partial \Sigma/\partial t = 0$, we only solve the dust evolution equations, but we distinguish two layers of dust with different radial diffusion coefficients $D_{\rm mid} = 0.033$ and  $D_{\rm sur} = 0.07$ from Table \ref{table:diffusion coeffs}. We further apply $\alpha = 0.02$ and make use of $D_z = 0.013$. The dust evolution is then solved using Equations \ref{eqn:dust-gas-fullcouple1} \& \ref{eqn:dust-gas-fullcouple2}, neglecting feedback on the gas profile, where initially a ring of dust with $T_S = 0.001$ is located at $R_0$.  The $F_z$ term represents vertical diffusion between two layers.
The time-independent gas density (black) and the radial advection velocities (red) are shown in Figure \ref{fig:steadystate-idealmhd}(c). To quantify the extent of outward grain diffusion in our MHD case and better compare two layers, instead of directly showing the radial evolution of $C$, which is much more quiescent than Figure \ref{fig:evolution-dust-gas-coupled-earlybump}(c) and Figure \ref{fig:evolution-dust-gas-coupled-grvmhd}(c), we compute the maximum fraction of dust mass that reaches the region $R>R_0$ beyond the initial radial location of the dust ring, defined as

\begin{eqnarray}
P_i(R) = \max_{0<t<T} \frac{\int_{R}^{\infty}2\pi R \Sigma C_i dR}{\int_{R_{\rm in}}^{\infty}2\pi R \Sigma C_i dR}
\end{eqnarray}
where $R_{\rm in}=1.0$ is taken as the position of the inner absorbing boundary, $T=1\times10^4~\Omega^{-1}$ the total simulation time, and $i={\rm mid, sur}$ for the midplane layer and surface layer respectively. 
Figure \ref{fig:steadystate-idealmhd}(a)(b) show the maximum fractions of dust mass reaching different radius $R>R_0$, 
for two typical values of $R_0=4$ and $R_0=10$. The insets of Figure \ref{fig:steadystate-idealmhd}(a)(b) show axisymmetric 2D distribution of $P(R)$ from the upstream fraction curves. In both cases, the surface layer display stronger upstream diffusion compared to the middle plane due to an order-unity larger $D_{\rm sur}$, although the midplane dust can also diffuse outwards significantly in the ideal MHD setup.


\subsection{Dust evolution in a steady-state MRI disk with non-ideal MHD turbulence}
\label{sec:MRI_with_dead_zone}

We then consider the effect of non-ideal MHD turbulence, where the midplane of the disk is much less ionized and less coupled to the MRI turbulence~\citep{gammie1996,kretke2010structure}, with effective $\alpha_{\rm mid} \sim 10^{-3}$. Hence, the diffusion in the middle plane and the vertical direction may be constrained by $\alpha_{\rm mid}$ and much less effective. We test this effect by going beyond simulation measurements and reducing coefficients to $D_{\rm mid}=D_z=0.003$. The midplane layer's $v_g$ and $v_d$ are also reduced according to a smaller $\alpha_{\rm mid}=0.003$. {In 3D non-ideal MHD simulations, gas in certain regions of the upper layer $|z|\gtrsim H$ can have positive radial velocity, albeit significant outflow ($v_g \sim v_K$) only occurs at $|z|\gg H$ where gas density is negligible \citep{Bai2017}. In our 2D treatment we still make the simplification to match a characteristic negative $v_g$ of the MRI-active layer with the inward accretion rate to ensure steady-state \citep{gammie1996}, proportional to the characteristic $\alpha$ that we assume in that layer. Note that if our upper layer has a significant positive gas velocity, dust outward transport will be aided by a combination of outward advection and diffusion, which will appear even more significant. Nevertheless, currently we only focus on the effect of the latter.}
The results are presented in Figure \ref{fig:steadystate-nonidealmhd} similar to Figure \ref{fig:steadystate-idealmhd}. The midplane now shows much smaller upstream fractions, as seen in (a)(b). { The gas inflow velocity is much larger near the surface
than the mid-plane, due to $\alpha_{\rm sur} \gg \alpha_{\rm mid}$, and the dust
inward-drift speed near the surface is also faster than that in the mid-plane (c).  
Nevertheless, upstream diffusion is much more pronounced near the surface than the 
mid-plane, due to $D_{\rm sur}\gg D_{\rm mid}$.}


Additionally, we consider the dust sizes in the midplane being larger (super-mm) than the surface layer (sub-mm or micron) due to relaxation of the fragmentation barrier (Equation \ref{eqn:frag}), producing a larger characteristic Stokes number for the midplane characteristic dust species $T_{\rm s,mid}=0.01$. Results from this setup is shown in Figure \ref{fig:steadystate-Ts}.
The difference in dust velocities for the two layers are significantly enhanced, as seen in Figure \ref{fig:steadystate-Ts}(c), while segregation effect becomes stronger and the surface layer can diffuse upstream significantly while the midplane is dominated by slow inward radial drift, as seen in Figure \ref{fig:steadystate-Ts} (a) and (b).
Changing the stopping time for mid-plane layer to $T_{\rm s,mid}=0.1$ could enhance this segregation effect even further.

\subsection{Dust evolution in an ideal MRI disk with stellar radiation wind}

At the final stages of gaseous disk evolution, accretion heating becomes subdominant and the central star illuminates the outer disk \citep{Nakamoto1994,Hueso2005}, which could cause significant evaporation of the gas density \citep{takeuchi2003}. To study the sole effect of radiation wind, based on the ideal MHD model (\S \ref{sec:idea_MHD}), we modify Equation \ref{eqn:gas density evo} to include a sink term
\begin{eqnarray}
\dfrac{\partial \Sigma}{\partial t}  = & \dfrac{3}{R} \dfrac{\partial}{\partial_R}\left( R^{1/2} \dfrac{\partial}{\partial_R} ( \Sigma \nu R^{1/2}) \right) - \dfrac{\Sigma}{\tau_{\rm wind}} f(R)
\label{eqn:gas density evo wind}
\end{eqnarray}
where $f(R)=\left(\tanh[(R-R_w)/\sigma_w]+1\right)/2$ is a smoothing function. 



We solve the modified gas evolution equation together with Equations \ref{eqn:dust-gas-fullcouple1} and \ref{eqn:dust-gas-fullcouple2}, choosing fiducial parameters within a range around $\tau_{\rm wind}\sim 100$, $R_w \sim 15$ and $\sigma_w \sim 5$. All other parameters are similar to that of Figure \ref{fig:steadystate-idealmhd}. 
The reduction of gas density beyond $R_w$ induces an outward gas velocity, on the same order of magnitude as the steady-state inward gas velocity. The resulting dust evolution is similar to the ideal MHD case presented in Figure \ref{fig:steadystate-idealmhd}, and both the surface layer and middle plane show considerable upstream diffusion.

\section{Summary}
\label{sec:summary}

Based on measurements from numerical simulations, we constructed 1D accretion disk models of gas and dust to determine outcomes of long-term diffusion process. In the simulations, we separate the disk into the midplane $|z|\lesssim H$ section and the surface section $|z|\gtrsim H$. Characteristic radial diffusion coefficients for each layer $D_{mid}$ and $D_{sur}$ are measured from Lagrangian particle diffusion, 
while exchange of particles between two layers is controlled by an overall vertical diffusion coefficient $D_z$. In the GI and GI-MHD runs, we found $D_z \sim \alpha \sim 0.1-0.2$ is quite efficient and the radial diffusion has little vertical dependence $D_{mid}\approx D_{\rm sur} \sim D_{\rm r}$, 
therefore in \S \ref{sec:GI_MHD} we only consider one homogeneous layer of $\mu$m size dust with characteristic Stokes number $T_s = 0.001$.
The early stage of a PSD is GI or GI-MHD dominated and highly dynamical, therefore we apply impulsive initial conditions with an evolving gas profile. 
We found very efficient diffusion of dust with $T_S\lesssim 0.001$ due to $D_{\rm r}$ much larger 
than the gas accretion coefficient $\alpha$. This implies that not only micron-size crystalline 
(highly refractory) grains condensed interior to the 
main belt of asteroid region can be transported outwards to Heliocentric distance of the 
Kuiper Belt region and be incorporated into comets (such as those retrieved from 
comet 81P Wild 2 by the Stardust mission), but up to mm-size CAI condensed 
in the sub-au inner regions are also prone to marginal gravitational instability, and can spread
out to the super-au main-belt regions to be assimilated into the planetesimal parent 
bodies of chondritic meteorites.

At a later stage, we consider a relatively quiescent MRI dominated disk with a fixed density profile; 
We first apply a diffusion model with two layers of dust of different diffusion coefficient $D_{\rm sur}>D_{\rm mid}$ according to the MHD simulation. Although diffusion coefficients are much smaller than in the  gravito-turbulence case, we still observe upstream diffusion of dust in both layers. That is to say, a considerable fraction of dust may reach the region beyond a its initial radial location through the effect of concentration diffusion. While the difference between two layers is small for our ideal-MHD model (\S \ref{sec:idea_MHD}), we also explore the influence of the existence of dead zone (\S \ref{sec:MRI_with_dead_zone}), where the turbulence parameter $\alpha$ is significantly smaller and characteristic dust Stokes number $T_{\rm S, mid}$ can be significantly larger than the surface layer $T_{\rm S, sur}$. In this case, the upstream diffusion of mm or cm grains in the dead zone
is much less vigorous than grains in the surface. {This effect 
attributes to the difference in disk radii inferred from NIR emission by mm-size dust in the midplane, or scattered light by micron-size grains
near the disk surface \citep{Villenave2020, Benisty2022}.} We lastly consider photo-evaporating winds in the outer regions of the disk, and found its sole effect can mildly enhance the upstream diffusion, albeit its efficiency is limited. 



\begin{acknowledgments}
We thank William Bethune, Hans Baehr, Zhaohuan Zhu and Wenrui Xu for 
useful conversations. We thank the anonymous referee for helpful suggestions. We have benefited from many past discussions with Willy Kley and would like to dedicate this paper
to his memory. Y. X. C. thanks Man Hoi Lee and Hong Kong University for their hospitality during the completion of this manuscript. T.Z. is supported by the Cecil and Sally Drinkward postdoc fellowship during this work. 
\end{acknowledgments}

%




\bibliography{main}{}
\bibliographystyle{aasjournal}



\end{document}